\UseRawInputEncoding
\documentclass[sigconf]{acmart}

\AtBeginDocument{%
  \providecommand\BibTeX{{%
    \normalfont B\kern-0.5em{\scshape i\kern-0.25em b}\kern-0.8em\TeX}}}

\setcopyright{acmcopyright}
\copyrightyear{2018}
\acmYear{2018}
\acmDOI{XXXXXXX.XXXXXXX}

\acmConference[ICAIF ’22]{Make sure to enter the correct
  conference title from your rights confirmation emai}{November 02–04, 2022}{New York, NY, USA}

\acmPrice{15.00}
\acmISBN{978-1-4503-XXXX-X/18/06}

\usepackage{algorithm}
\usepackage{algorithmicx, algpseudocode}
\usepackage{graphicx}
\usepackage{multirow}

\theoremstyle{definition}
\newtheorem{definition}{Definition}

\begin{document}

\title{Reinforcement Learning Portfolio Manager Framework with Monte Carlo Simulation}


\author{Jungyu, Ahn}
\authornote{Corresponding author.}
\affiliation{%
  \institution{Shinhan AI}
  \country{Seoul, Republic of Korea}
}
\affiliation{%
  \institution{Department of Computer Science and Engineering}
  \city{Inha University}
  \country{Incheon, Republic of Korea}
}
\email{ahnjungyu320@gmail.com}

\author{Sungwoo, Park}
\affiliation{%
  \institution{Shinhan AI}
  \country{Seoul, Republic of Korea}
}
\email{swpark723@gmail.com}

\author{Jiwoon, Kim}
\affiliation{%
  \institution{Shinhan AI}
  \country{Seoul, Republic of Korea}
}
\email{jiwoonkim83@gmail.com}

\author{Ju-hong, Lee}
\affiliation{%
  \institution{Department of Computer Science and Engineering}
  \city{Inha University}
  \country{Incheon, Republic of Korea}
}
\email{juhong@inha.ac.kr}

\renewcommand{\shortauthors}{Jungyu, Ahn et al.}

\begin{abstract}
Asset allocation using reinforcement learning has advantages such as flexibility in goal setting and utilization of various information. However, existing asset allocation methods do not consider the following viewpoints in solving the asset allocation problem. First, State design without considering portfolio management and financial market characteristics. Second, Model Overfitting. Third, Model training design without considering the statistical structure of financial time series data. To solve the problem of the existing asset allocation method using reinforcement learning, we propose a new reinforcement learning asset allocation method. First, the state of the portfolio managed by the model is considered as the state of the reinforcement learning agent. Second, Monte Carlo simulation data are used to increase training data complexity to prevent model overfitting. These data can have different patterns, which can increase the complexity of the data. Third, Monte Carlo simulation data are created considering various statistical structures of financial markets. We define the statistical structure of the financial market as the correlation matrix of the assets constituting the financial market. We show experimentally that our method outperforms the benchmark at several test intervals.
\end{abstract}

\ccsdesc[500]{Information systems~Data stream mining}

\keywords{Reinforcement Learning, Asset Allocation, Financial Time Series, Monte Carlo Simulation}


\maketitle

\section{Introduction}
Investment refers to allocating time and money in anticipation of future profits.\cite{bodie2020investments}. There are many different ways to allocate time and money.For example, there is a method of depositing an investment in a bank, a method of investing in a stock-type fund, a method of investing in a bond-type fund. These methods are either low- or high-returns methods; however, they are prone to market risks and may result in a loss of principal. It is important to adjust the ratio of risky and safe assets to avoid low-return or high-risk situations. To classify risky and safe assets, they should first be categorized into asset classes according to the following characteristics. Assets within an asset class should be homogeneous (Homogeneity); no overlap should exist between asset classes (exclusiveness); and a diversification effect should exist when investing using asset class (Diversification) \cite{bodie2020investments},\cite{gibson2008asset}. Using these characteristics, assets should be categorized into safe asset class and risky asset class. And a portfolio developed with these assets should be managed with low risk and high expected return. Asset allocation allocates money to different assets with different expected returns and different levels of risk. Thus, a portfolio is developed for achieving low volatility and long-term stable returns. \cite{bodie2020investments}, \cite{meucci2005risk}. As a method of asset allocation, there is a traditional asset allocation method, and recently, an asset allocation method using machine learning is being developed.

There are many different method for traditional asset allocation, such as Markowitz and Risk Budgeting. Markowitz was proposed by Harry Markowitz. In this model, the historical prices of assets are used to estimate the expected return and risk of a portfolio. and these statistics are used to derive efficient frontier, which are subsequently used to derive an optimal portfolio. \cite{markowitz1952portfolio}. Risk budgeting estimates portfolio weights by assigning a portfolio's risk to each asset. When allocating risks to an asset, the risk tolerance of the investor should be considered for the asset.\cite{pearson2011risk}. Recently, financial AI techniques using machine learning are being actively studied. The results of these studies are being used in areas such as market forecasting, asset allocation, risk management, and market timing. \cite{tran2021data}, \cite{uysal2021machine}, \cite{mascio2021market}, \cite{ma2021portfolio}. reinforcement learning is a method of learning how an agent recognizes the current state in an environment and selects an action to maximize reward.\cite{sutton2018reinforcement}. Asset allocation using reinforcement learning has more flexibility in setting goals and using input variables compared to the existing asset allocation model. For example, asset allocation using reinforcement learning can take into account a variety of information as a financial market environment. Such a variety of information includes benchmark management information, portfolio management information, market returns, market price prediction information, market volatility information, basic information of stocks, macro variables. Such a variety of information cannot be directly considered in the existing asset allocation method.

Asset allocation using reinforcement learning has advantages such as flexibility in goal setting and utilization of various information. However, the existing asset allocation methods do not consider the following viewpoints in solving the asset allocation problem.

\begin{enumerate}

\item \textbf{State design without considering portfolio management and financial market characteristics}\\
Actions computed by the model should not change the financial markets, but change the portfolio operated by the agent. Investments by institutions or pension funds can bring big changes to the financial market, but individual investments cannot bring big changes to the financial market. However, the existing asset allocation methods using reinforcement learning only consider changes in the financial market as the agent's state.This method can be interpreted as changing the state of the financial market rather than changing the state of the managed portfolio as a result of investment.

\item \textbf{Model Overfitting}\\
Assuming that 20 years of financial time series daily data is used to train the model, one stock or ETF has approximately 4800 daily data. Machine learning, reinforcement learning method use historical financial time series data as training, validation and test. In this case, 60$\%$ of the total data are used as training data, 30$\%$ data are used as validation data, and 10$\%$ are used as test data. In this data split, deep learning and machine learning architectures can have high model complexity compared to data complexity. Therefore, the probability that the model overfits the training data is very high.\cite{murphy2012machine}, \cite{goodfellow2016deep}. An overfitted model can perfectly manage the portfolio for training data, but the model does not manage the portfolio well for validation and test data, which can lead to poor portfolio performance.

\item \textbf{Model training design without considering the statistical structure of financial time series data}\\
Because financial time series data has non-stationary characteristics, the statistical structure of financial time series data is constantly changing.\cite{xue2018financial},\cite{therrien2018probability},\cite{tran2021data}. However, the existing asset allocation method using reinforcement learning does not consider the various statistical structures of financial time series data in model learning. In these methods, the various statistical structures in the training data are averaged.
\end{enumerate}

To solve the problem of the existing asset allocation method using reinforcement learning, we propose a new reinforcement learning asset allocation method. First, the state of the portfolio managed by the model is considered as the state of the reinforcement learning agent, and the daily return vector of the asset is considered as observation. In other words, the agent takes into account changes in financial markets and changes in the portfolio managed by the agent simultaneously to maximize rewards. Second, we increase the complexity of the training data using Monte Carlo simulation data to prevent model overfitting. These data can have different patterns, which can increase the complexity of the data.This can alleviate the problem of model overfitting, a chronic problem with financial time series data. Third, we create Monte Carlo simulation data in consideration of various statistical structures of financial markets. We define the statistical structure of the financial market as the correlation matrix of the assets constituting the financial market. The various statistical structures of financial markets are considered with k different correlation matrices.Using these correlation matrices and the Black-Scholes Merton algorithm \cite{black2019pricing}, we generate a monte carlo simulation data set.

This paper is organized as follows. Section 2 describes related work, Section 3 describes motivation, Section 4 describes proposed method, Section 5 present the experiments, Section 6 describes conclusion and future work.

\section{RelatedWork}
Traditional asset allocation method are still being actively studied and applied to finance. \cite{ji2018risk} proposed a new probabilistic risk budgeting multi-portfolio optimization. This model performs an integrated optimization of the individual security responsible for marginal risk contribution. \cite{gabler2019extension} applied the Markowitz methodology to develop a new strategy algorithm for long-term stocks. This methodology optimizes portfolios based on high potential returns and high risk strategies.

There are various method for asset allocation using reinforcement learning. \cite{noguer2020deep} applied a simple deep reinforcement learning method to US Equity. \cite{hu2019deep} built an asset allocation model using a GRU network and a deep reinforcement learning (DRL) algorithm. This model uses previous state, action to extract the current action. And the model was trained using risk-adjusted returns.  \cite{pendharkar2018trading} used the S$\&$P 500 and U.S. Treasury bonds as assets, used adaptive learning, and built a retirement pension trading portfolio model. \cite{ye2020reinforcement} used financial article data and stock price data to generate stock movement signals, and used the signals and stock price data to build an asset allocation reinforcement learning agent.

 \cite{benhamou2021detecting} developed a portfolio reinforcement learning agent using three subnetworks. The first subnetwork is a network that handles returns and volatility of assets, and the second subnetwork is a network that handles contextual information such as correlations between assets. The third subnetwork handles the previous portfolio weights.

 \section{Motivation}
 \subsection{Monte Carlo Simulation}
Monte Carlo simulation is a numerical method that can be effectively applied to problems with unknown or complex closed solutions.\cite{robert2004monte} Monte Carlo simulations are being used in various fields such as stock prices and interest rates that follow a probabilistic model. Monte Carlo simulations are used in the valuation of derivatives. The stochastic process mainly used for the movement of financial assets is the geometric Brownian motion. A stochastic process $S_{t}$ is said to follow geometric Brownian motion when it satisfies the following stochastic differential equation:
\begin{equation}
dS_{t}=\mu S_{t} dt + \sigma S_{t} dW_{t}
\end{equation}

In this case, $\mu$ is the return of the underlying asset, $\sigma$ is the volatility of the return, and $W_{t}$ is the winner process. When the initial value is $S_{0}$, the solution $S_{t}$ in the above equation satisfies the following equation.
\begin{equation}
S_{t}=S_{0} exp [ (\mu - \frac{1}{2} \sigma^{2} ) t + \sigma W_{t} ]
\end{equation}

We can also derive the following equation to run the Monte Carlo simulation. $\Delta t$ is the time interval and $\epsilon_{t}$ is a standard normal random number. Using the above formula, we can create multiple stock paths. And we use multiple stock paths to calculate the payoff for the option and determine the option price.

\begin{equation}
S_{t+ \Delta t } = S_{t} exp [(\mu - \frac{1}{2} \sigma^{2}) \Delta t + \sigma \epsilon_{t} \sqrt{\Delta t}
\end{equation}

In addition, the stock price path of assets can be created using correlation of assets and Monte Carlo simulation. These stock paths are paths that satisfy special correlations between assets. And thousands or tens of thousands of asset paths can be created through a specific correlation matrix. And these asset paths have various patterns that can be expressed in real financial markets.

\subsection{Changes in The Financial Market Statistical Structure}
Financial time series data are generally non-stationary.\cite{xue2018financial},\cite{therrien2018probability}. This means that the mean and variance of the random process that generates financial time series data are constantly changing.

\begin{figure}[h] 
\begin{center}
\includegraphics[width=1.0\linewidth]{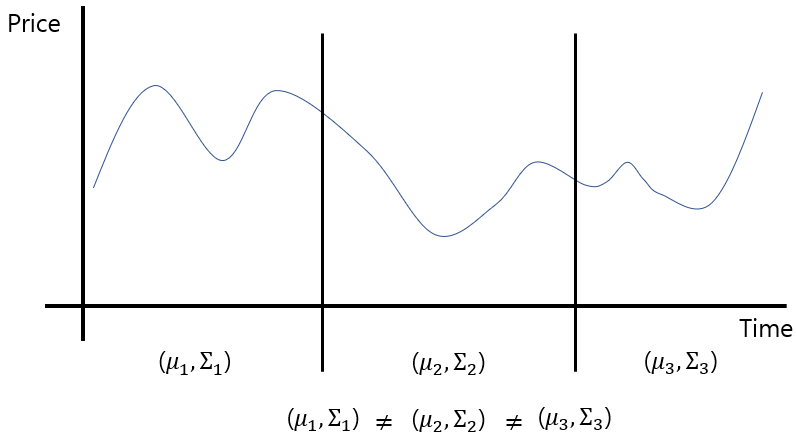}
\end{center}
\caption{Changes in mean and variance}
\end{figure}

If the statistics of the random process that generate financial time series data change, the model should be optimized for the changed financial time series statistics. However, these characteristics are not considered in the existing asset allocation method using general machine learning and reinforcement learning. the various statistical structures of financial markets are not considered. It simply learns only the statistical structure of the financial market included in the training data.

\begin{figure}[h] 
\begin{center}
\includegraphics[width=1.0\linewidth]{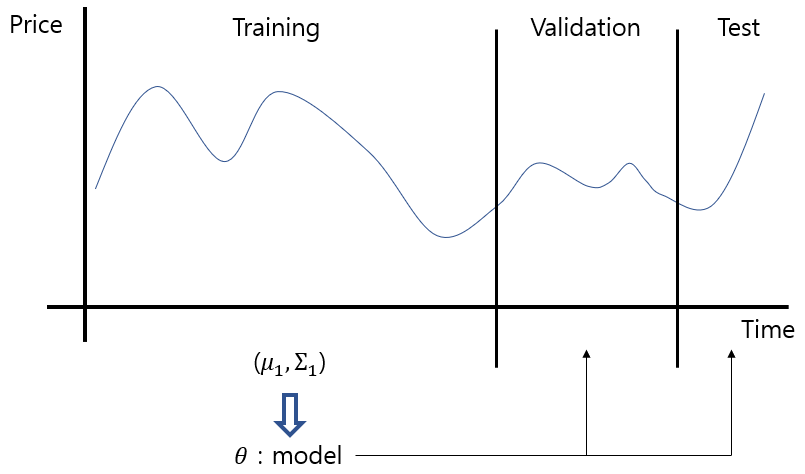}
\end{center}
\caption{general model training and evaluation method}
\end{figure}

This has the disadvantage that the model does not take into account the various statistical structures of the financial market. Therefore, in this paper, we create an asset allocation model that considers the various statistical structures of the financial market.

\subsection{Model training using simulation data}

For financial time series data, the size of the training data set is usually small. Therefore, it is difficult for reinforcement learning models to learn patterns of various financial time series data. Reinforcement learning models trained on a small number of data may suffer from increased losses during testing or asset management.  Therefore, in order to train a reinforcement learning model, it is important to have training data of various patterns. Training data with various patterns can increase the complexity of the training data and prevent the model from overfitting. And preventing model overfitting can prevent an increase in loss during model testing or asset management.

\section{Proposed Method}
 In this section, we define the financial time series simulation data generation method, model architecture and state, reward, learning method, and test method. The following is a reinforcement learning portfolio manager framework.

\begin{figure}[h] 
\begin{center}
\includegraphics[width=1.0\linewidth]{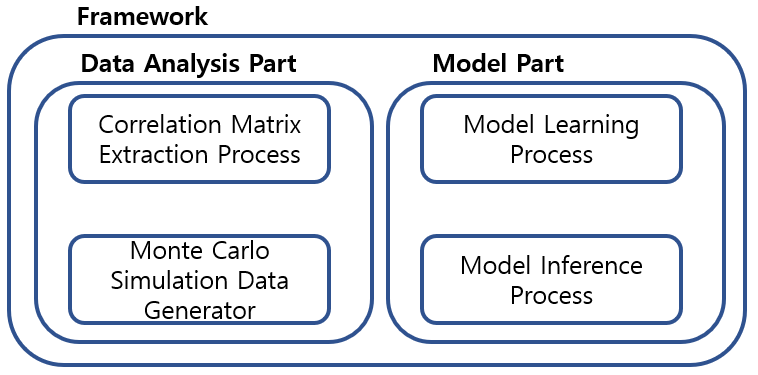}
\end{center}
\caption{Reinforcement Learning Portfolio Manager Framework}
\end{figure}

\subsection{Reinforcement Learning Portfolio Manager Framework}
The reinforcement learning portfolio manager framework is largely categorized into a data analysis part and a model part. The data analysis part consists of two processes. First, the process of extracting the statistical structure of financial markets. Second, the process of generating financial time series simulation data using the statistical structure. The model part also consists of two processes.First, the process of learning the model using financial time series simulation data, second, the process of inferring the output of the model with respect to the test data

\subsection{Data Analysis Part}
\subsubsection{\textbf{Creation of financial time series simulation data}}
In this paper, model training is performed using financial time series simulation data. To generate financial time series simulation data, we use Black-Scholes-Merton method\cite{black2019pricing}. The Black-Scholes-Merton method can generate various patterns of financial time series simulation data considering the statistical structure of the financial market. The statistical structure of the financial market is regarded as the correlation matrix of the assets constituting the financial market. Financial time-series simulation data is generated in two steps. First, extraction of representative correlation matrix to know the statistical structure of financial market, second, generation of financial time series simulation data. Financial time series simulation data generation can be represented by the following figure(data analysis part).

\begin{figure}[h] 
\begin{center}
\includegraphics[width=1.0\linewidth]{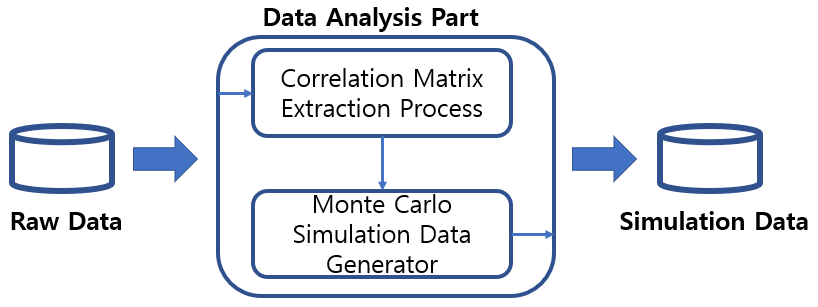}
\end{center}
\caption{data analysis part}
\end{figure}

To extract the representative correlation matrix, we define the Representative Correlation Matrix Extraction Process (RCME Process). The RCME process divides financial time series data into several specific financial time series intervals and extracts a representative correlation matrix from the intervals. Each representative correlation matrix extracted from each interval is different. A representative correlation matrix extracted from a specific time interval models the linear relationship of assets that exist in that time interval. The detailed RCME Process is as follows. The correlation matrix of the set of assets at each time t of the financial time-series data is computed. And the process creates a correlation matrix set (CMS) using these correlation matrices. The size of the CMS is T, where T is the total length of the financial time series. The process calculates the pairwise distance using the correlation matrices present in the CMS. When the process completes calculating the distances between the correlation matrices in the CMS, a correlation matrix distance matrix (CMDM) is generated. Each element of CDMD is defined as follows.

\begin{definition}[Correlation Matrix Distance Matrix Element $D_{i,j}$]
\begin{equation}
D_{i,j}= \| A^{i,j} \|_{F}, i,j=1,...,T
\end{equation}
\end{definition}

\begin{definition}[Frobenious norm with two correlation matrix]
\begin{equation}
\| A^{i,j} \|_{F} = \sqrt{\sum_{k=1}^{n}\sum_{l=1}^{n} |a_{k,l}|},
\end{equation}

Where :$ A^{i,j}=\Sigma_{i}-\Sigma_{j}, \Sigma_{i}, \Sigma_{j} \in CMS$

\end{definition}

After the CMDM is created, the Process executes Hierarchical Clustering\cite{han2011data} to create a cluster set (CS). 

For each cluster generated by hierarchical clustering, the process computes a representative correlation matrix. The representative correlation matrix is calculated as the average matrix of the correlation matrices belonging to the cluster.

\begin{definition}[Representative Correlation Matrix]
Representative correlation matrix for a specific cluster
\begin{equation}
\overline{\Sigma}^{i}=\frac{1}{k} \sum_{j}^{k} \Sigma_{j}^{i}
\end{equation}

Where : $\Sigma_{j}^{i}$ jth correlation matrix in ith cluster

\end{definition}

The representative correlation matrix is the representative correlation matrix representing the correlation matrices belonging to the cluster. And correlation matrices belonging to the same cluster are correlation matrices extracted from similar financial time series data. The representative correlation matrix is a correlation matrix that represents the cluster. This matrix can be considered to be a correlation matrix representing a particular interval of financial time-series data.

\begin{algorithm}[h]
\caption{representative correlation matrix extraction process}
\begin{algorithmic}[1]
\Require Financial Time Series Data Set
\Ensure Representative Correlation Matrix Set
\For{i $\leftarrow$ 1 to T}
\State $\Sigma_{i} \leftarrow $ calculate correlation matrix in i th time
\State CMS $\leftarrow$ $\Sigma_{i}$
\EndFor

\For{i $\leftarrow$ 1 to T}
\For{j $\leftarrow$ 1 to T}
\State $D_{i,j} \leftarrow \| A^{i,j} \|_{F}$
\EndFor
\EndFor

\State CS $\leftarrow $ Hierarchical Clustering(CMDM)

\For{i $\leftarrow$ 1 to K}
\State $\overline{\Sigma}^{i} \leftarrow$ average correlation matrices in i th cluster
\EndFor

\end{algorithmic}
\end{algorithm}

We generate financial time series simulation data using the representative correlation matrix calculated by the RCME process. We use the Black-Scholes-Merton method to generate the financial time series simulation data\cite{black2019pricing}. The financial time series simulation data generated by Black-Scholes-Merton \cite{black2019pricing} are various simulation data corresponding to the correlation matrix, the statistical structure of the financial market. The data includes various patterns. The actual financial time series data is only one realization path data realized by the correlation matrix, and only specific patterns are included in the data. In machine learning methods using financial time series data, these various patterns can increase data complexity and prevent model overfitting.

\subsection{Agent Action, Observation, State, Reward}
\subsubsection{\textbf{Agent Action Set Extraction}}

Action set extraction is the process of defining an agent's actions. this process is defined using the mean and variance of portfolio management returns.

First, we define up/down intervals in financial time series data. The up/down point is defined as the point in time when the average of daily returns over the past k days from a point in time t is above $\alpha \% $ or below $- \alpha \% $. A up/down point is defined for each point in time in the financial time series data. these successive up/down points are used to define the up/down interval. For example, if the period 2005 to 2008 consists of up points, we define that period as an up interval. if the interval 2008 to 2009 consists of a down points, we define that period as a down interval. 

Second, the agent action set is extracted using the up/down intervals. An agent action set is a set of actions that agents can select from. In order to define the agent action set, the investment weight set and the asset allocation weight vector set must be defined first. An investment weight set is a set of asset allocation weights that can be assigned to each asset. For example, if the investment weight set is defined as $[$ 0 $\%$, 10$\%$, 20 $\%$ $]$, each asset is invested with one of 0$\%$, 10$\%$, 20$\%$. We define a set of asset allocation weight vectors using the investment weight set. Cartesian products are performed using the investment weight set. From the set of vectors resulting from the Cartesian product, only vectors whose sum of vector elements are 1 are extracted. A set of asset allocation weight vectors is created using the extracted vectors. We extract the action set of the agent in the up/down intervals using the asset allocation weight vector set.The detailed steps to extract an action are as follows. One interval is selected in the up/down section. A random sampling is then performed on the set of asset allocation weight vectors. The process uses the sampled asset allocation weight vector to calculate portfolio management returns for the selected interval. the process uses portfolio management returns to calculate the mean return ($\mu$) and volatility ($\sigma$) of a portfolio. The process calculates an action evaluation measure using the calculated mean return and volatility of the portfolio.

\begin{definition}[Action Evaluation Measure]
\begin{equation}
f(\mu,\sigma,k)= \mu -k \times \sigma
\end{equation}

Where : $k$ is control term
\end{definition}

The reason for random sampling is as follows.If the size of the set of asset allocation weight vectors is very large, it is impossible to construct a space for Action Evaluation Measure scale for the set. Therefore, it is much more efficient in terms of time efficiency to approximate the action evaluation measure space by randomly sampling the asset allocation weight vector from the set. Experimentally, the statistic of the action evaluation measure in the entire set of asset allocation weight vectors is maintained in the sample set of asset allocation weight vectors. The size of the sample set is 0.01$\%$ of the total set. In the approximate space of a one-dimensional action evaluation measure, the process selects i actions with large values and includes them in the agent action set. By applying the process to all up/down intervals, the process constitutes an agent action set.

\subsubsection{\textbf{observation, state, reward}}
 An agent's observation is defined as a return matrix consisting of past 60 day return vectors based on time t for individual assets. The definition of the return vector and return matrix is as follows.

\begin{definition}[Historical Return Vector]-Observation
\begin{equation}
\textbf{r}_{t}=(r_{t-59}, r_{t-58},...,r_{t})
\end{equation}

Where : $r_{t}=\frac{asset price_{t}}{asset price_{t-1}}-1$
\end{definition}

\begin{definition}[Historical Return Matrix]-Observation
\begin{equation}
\bar{\bar{\textbf{r}}}_{t}= \begin{bmatrix}
\textbf{r}_{t}^{1} \\ ... \\ \textbf{r}_{t}^{k}
\end{bmatrix}
\end{equation}

where : $\textbf{r}_{t}^{i}$ is the historical return vector of the ith asset at time t
\end{definition}

The state of an agent is defined as the return vector for past 120 days of the portfolio managed by the agent at time t.  

\begin{definition}[Historical Portfolio Management Return]-State
\begin{equation}
\textbf{pr}_{t}=(pr_{t-119}, pr_{t-118},...,pr_{t})
\end{equation}

Where : $r_{t}=\frac{portfolio value_{t}}{portfoliovalue_{t-1}}-1$
\end{definition}

In general, reinforcement learning agents in portfolio management in other method tend to use asset return as the agent's state. However, this information is not information based on the agent's actions. This is simply information about the market and not information that depends on an agent's actions. Therefore, it is reasonable to define the return of the portfolio depending on the agent's action as the agent's state. The reward is defined as the sharpe ratio of portfolio management performance. The sharpe ratio is as follows.

\begin{definition}[Sharpe Ratio]-Reward
\begin{equation}
Sharpe Ratio=\frac{\mu_{port}}{\sigma_{port}}
\end{equation}

\end{definition}

$\mu_{port}$ is the average value of the daily returns of the portfolio managed by the agent. And the value is annualized. $\sigma_{port}$ is the volatility of the daily return of the portfolio managed by the agent. And the value is annualized. In order to maximize the Sharpe ratio value, $\mu_{port}$ should be maximized and $\sigma_{port}$ should be minimized. We set the sharpe ratio as the agent's reward, and the agent learns to keep the portfolio management performance upward. 

\subsection{Model Learning}
The model is trained through the following process. The learning process selects one of the correlation matrices generated by the RCME process. The simulation data set is generated by applying the Black-Scholes Merton \cite{black2019pricing} method to the selected correlation matrix. The generated simulation data set is paired with the selected correlation matrix. The learning process trains a reinforcement learning model using a simulation data set.

As the input of the reinforcement learning model, the Historical Return Matrix is used for Observation, and the Historical Portfolio Management Return is used for the State. The features of observation are extracted by applying multiple convolution filters to observation. And the features of the state are extracted by applying multiple convolution filters to the state.And the nonlinear relationship is modeled by applying the features extracted from observation and state to the FC layer. The result of FC Layer is used as input for value function and policy function, respectively. A3C is used as the architecture learning algorithm. The models trained for the selected correlation matrix are defined as a sub model pool. The total model pool is defined as sub model pools. The sub model pool and the total model pool are represented by the following figure.

\begin{figure}[h] 
\begin{center}
\includegraphics[width=0.8\linewidth]{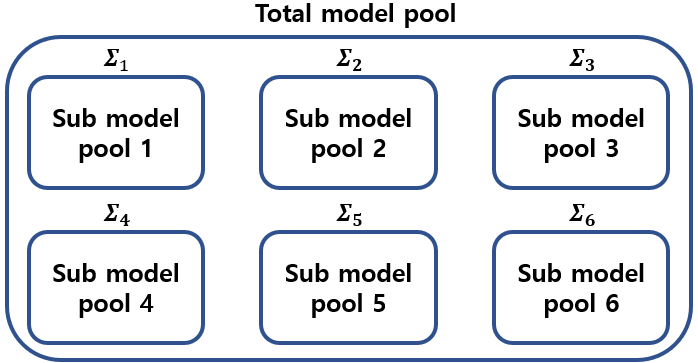}
\end{center}
\caption{Total Model Pool, Sub Model Pool}
\end{figure}

Model inference is performed as follows. The inference process chooses a representative correlation matrix that is most similar to the correlation matrix computed using historical data at time t. Inference is performed at time t using the sub-model pool associated with the corresponding representative correlation matrix.

\begin{algorithm}[h]
\caption{Model Learning Process}
\begin{algorithmic}[1]
\Require Representative Correlation Matrix Set
\Ensure Total Model Pool
\State T: Total Model Pool
\State S: Sub Model Pool
\State RCMS : Representative Correlation Matrix Set
\State BSM : Black Scholes Merton
\State SDS : Simulation Data Set

\For { \textbf{each} $\Sigma$ in RCMS}
\State SDS $\leftarrow$ BSM($\Sigma$)

\For{ \textbf{each} simulation data in SDS}
\State $\theta \leftarrow $ A3C(simulation data)
\State S $\leftarrow$ $\theta$
\EndFor

\State T $\leftarrow$ S
\EndFor
\end{algorithmic}

\end{algorithm}

\section{Experiment}

\subsection{Experiment Data}
To proceed with global asset allocation, the experimental data are an Exchange Traded Fund (ETF) that tracks stock indices, bonds, and commodities in major countries. The ETF's name is Bloomberg ticker.

\begin{table}[h]
\centering
\caption{Asset Index, Exchange Traded Fund(ETF)}
\begin{tabular}{cc|ccc}
\noalign{\smallskip}\noalign{\smallskip}\hline\hline
\multicolumn{2}{c|}{} & Index & ETF \\
\hline
\multirow{1}{*}{Equity} 
& USA  & S$\&$P 500 & VOO US Equity \\
& UK  & FTSE 100 & ISF LN Equity \\
& GERMANY  & DAX & XDAX GR Equity \\
& JAPAN  & NIKKEI & 1329 JP Equity \\
& CHINA  & FTSE CHINA 50 & FXI US Equity \\
& KOREA  & KOSPI & 277630 KS Equity \\

\hline
\multirow{1}{*}{Bond} 
& USA  & US 10Yr & IEF US Equity \\
& UK  & UK 10 Yr & GLTY LN Equity \\
& GERMANY  & GR 10 Yr & RXP5EX GR Equity \\
& JAPAN  & JP 10 Yr & 2510 JP Equity \\
& CHINA  & CH 10 Yr & 2813 HK Equity \\
& KOREA  & KR 10 Yr & 114260 KS Equity \\

\hline
\multirow{1}{*}{Commodity} 
& Gold  & Gold Trust & GLD US Equity \\

\hline
\hline
\end{tabular}
\end{table}

\subsection{Benchmark}
To evaluate the proposed framework, we use traditional asset allocation models and equal weight models as benchmarks. The traditional asset allocation model is a model widely used in the field of asset management. The Markowitz and Risk Budgeting model is used as a traditional asset allocation model. The risk budgeting model uses a risk parity model that equally allocates risk to each asset. Equal weight model is a model that allocates the ratio of asset allocation to each investment asset as 1/n. 

\subsection{Experiment Result}
We perform the experiment on 2008 $\sim$ 2021. Models in the proposed framework are trained using only simulation data generated by black-scholes merton \cite{black2019pricing}. For models in our framework, all experimental data is unseen and test data. The portfolio managed by the model is managed for two years. We set annualized return, annualized volatility, Sharpe's index, MDD, and cumulative return as statistics for evaluation.

\begin{definition}[Annualized Return]
\begin{equation}
r_{y}=\bar{r} \times 252
\end{equation}

Where : $\bar{r}=\frac{1}{n} \sum_{i}^{n} r_{i}$ ,$r_{i}$ is the daily return of the portfolio. n is the portfolio management period.

\end{definition}

\begin{definition}[Annualized Volatility]
\begin{equation}
\sigma_{y} = \bar{\sigma} \times \sqrt{252}
\end{equation}

Where : $\bar{\sigma}=\frac{1}{n-1} \sum_{i}^{n} (r_{i} - \bar{r} )^{2} $ 

\end{definition}

MDD refers to the maximum decline from the previous high during portfolio management.

\subsubsection{\textbf{Experiment Period Setup}}

Experiment Period is set as follows. (1) 2008-02-18 $\sim$ 2010-02-18, (2)2010-02- 18 $\sim$ 2012-02-18, (3) 2012-02-18 $\sim$ 2014-02-18,(4) 2014-02-18 $\sim$ 2016-02-18, (5)2016-02-18 $\sim$ 2018-02-18, (6) 2018-02-18 $\sim$ 2019-02-18. For Not Daily Rolling, each Period is an portfolio management period. For Daily Rolling, all internal dates of the period are the starting points of portfolio management.

\subsubsection{\textbf{Not Daily Rolling Result}}
the experiment start dates are February 18, 2008, February 18, 2010, February 20, 2012, February 18, 2014, February 18, 2016, February 19, 2018. , September 12, 2019. The experiment start dates are the start date of portfolio management, and the portfolio management period is 2 years. 

The Markowitz \cite{markowitz1952portfolio} model is characterized by a strong concentration of asset allocations on specific assets. In particular, this concentration of asset allocation is noticeable in stocks and gold. Stocks and gold are characterized by high asset returns but are highly volatile. Therefore, Markowitz \cite{markowitz1952portfolio} has the lowest sharpe value among the four models. Also, the absolute value of MDD is the largest among the four models. This can be the biggest decline from the peak when the asset is managed as a Markowitz \cite{markowitz1952portfolio}. this result indicates that the management performance is not stable.

In the case of risk budget \cite{pearson2011risk}, the proportion of asset allocation is focused on bonds. In this model, risk is controlled by adjusting the weight of bonds. This characteristic can be confirmed in that the average volatility value of the intervals is significantly lower than that of other benchmarks. However, it can be seen that the average annual return is the lowest among the four models. And it can be seen that the sharpe value is higher than other benchmarks due to the very low volatility. However, because the average return is very low, asset management profit is expected to be very low. In particular, the Risk Budgeting \cite{pearson2011risk} model is not suitable for a strategy that pursues a return. 

For equal weights, the average annual return for that period is lower than the Markowitz \cite{markowitz1952portfolio}, but the average annual volatility is much lower. Compared with Markowitz \cite{markowitz1952portfolio}, it can be seen that the asset management performance of the equal weight is not bad. Equal weight also have greater average annualized volatility for each interval than Risk Budgeting \cite{pearson2011risk}, but higher average annualized returns. Equal Weight is more suitable than risk budgeting model in terms of return. 

It can be seen that our model shows good returns and volatility compared to the benchmark algorithm. In particular, for sharpe value, which means stable operating performance, our model does not increase sharpe value by extremely low volatility, as in risk budgeting \cite{pearson2011risk}. Our model raises the sharpe value by operating a strategy to reduce volatility and bring moderate returns. Also, our model has a lower absolute MDD value for equal weight. This can be interpreted as our model is good at defending against a decline in the market compared to equal weight. In ALL Asset Class and DM Asset Class, it can be seen that our model is performing well compared to benchmarks. It can be seen that the performance is not limited to a specific asset class, but is observed in a variety of asset classes.

\begin{table*}[h]
\centering
\addtolength{\leftskip} {-2cm}
\addtolength{\rightskip}{-2cm}

\caption{Experiment Result, Not Rolling}
\label{t5}

\resizebox{16cm}{!}{
\begin{tabular}{|c|c|c|c|c|c|c|c|c|c|c|c|c|c|c|c|c|}
\hline\hline

 &
\multicolumn{4}{|c|}{Markowitz} &
\multicolumn{4}{|c|}{Risk Budgeting} &
\multicolumn{4}{|c|}{Equal Weight} &
\multicolumn{4}{|c|}{Model} \\

\hline
Period & R & Sigma & MDD & Sharpe & R & Sigma & MDD & Sharpe & R & Sigma & MDD & Sharpe & R & Sigma & MDD & Sharpe \\

\hline
(1) 
& -8.97 $\%$ & 59.70$\%$ & -64.95$\%$ & -0.05 
& 0.62$\%$ & 4.74$\%$ & -8.27$\%$ & 0.13 
& 1.93$\%$ &  -11.45$\%$ & -22.69$\%$ & 0.16 
& 2.68$\%$ & 7.18$\%$ & -11.74$\%$ &  0.36 \\

\hline
(2) 
& 15.27$\%$ & 20.78$\%$ & -24.14$\%$ & 0.73 
& 4.45$\%$ & 2.77$\%$ & -2.41$\%$ & 1.6 
& 6.77$\%$ &  6.92$\%$ & -9.37$\%$ & 0.97 
& 4.06$\%$ & 4.92$\%$ & -4.16$\%$ &  0.82 \\

\hline
(3) 
& 8.97$\%$ & 13.46$\%$ & -14.72$\%$ & 0.66 
& 0.90$\%$ & 2.42$\%$ & -4.86$\%$ & 0.37 
& 2.26$\%$ &  5.34$\%$ & -6.99$\%$ & 0.42 
& 1.49$\%$ & 4.39$\%$ & -5.81$\%$ &  0.34 \\

\hline
(4) 
& 4.67$\%$ & 19.95$\%$ & -27.55$\%$ & 0.23 
& 2.59$\%$ & 2.09$\%$ & -4.66$\%$ & 1.23 
& 0.76$\%$ &  5.35$\%$ & -12.13$\%$ & 0.14 
& 0.25$\%$ & 3.89$\%$ & -7.52$\%$ &  0.06 \\

\hline
(5) 
& 17.38$\%$ & 17.96$\%$ & -15.26$\%$ & 0.96 
& 1.12$\%$ & 1.85$\%$ & -3.49$\%$ & 0.6 
& 6.33$\%$ &  4.36$\%$ & -5.04$\%$ & 1.45 
& 4.38$\%$ & 3.62$\%$ & -4.72$\%$ &  1.21 \\

\hline
(6) 
& 9.26$\%$ & 14.90$\%$ & -21.72$\%$ & 0.62 
& 2.68$\%$ & 2.16$\%$ & -3.98$\%$ & 1.24 
& 3.18$\%$ &  4.62$\%$ & -7.54$\%$ & 0.68 
& 3.96$\%$ & 3.65$\%$ & -4.74$\%$ &  1.08 \\

\hline
(7) 
& 18.86$\%$ & 24.81$\%$ & -32.26$\%$ & 0.76 
& 1.80$\%$ & 3.48$\%$ & -7.20$\%$ & 0.51 
& 7.24$\%$ &  8.14$\%$ & -15.04$\%$ & 0.88 
& 7.85$\%$ & 7.09$\%$ & -11.88$\%$ &  1.1 \\

\hline
Mean 
& 9.35$\%$ & 24.51$\%$ & -28.66$\%$ & 0.38 
& 2.02$\%$ & 2.79$\%$ & -4.98$\%$ & 0.72 
& 4.07$\%$ &  6.60$\%$ & -11.26$\%$ & 0.61 
& 3.52$\%$ & 4.96$\%$ & -7.22$\%$ & 0.71 \\

\hline
\hline
\end{tabular}
}
\end{table*}

\subsubsection{\textbf{Daily Rolling Result}}
In addition to conducting experiments by setting a fixed operating period, a daily rolling experiment is required to measure the asset management performance of the model by changing the starting point of daily operation. Changing the starting point of operation every day has the following characteristics: Various financial time series patterns can be used as input to the model. It can be seen that the model works well for various financial time series patterns. In addition, it can be seen that portfolio management performance can be very different depending on the start time of management and the time of portfolio rebalancing. As a result of Daily Rolling, there is a characteristic that can present the average return and volatility of the fund to customers regardless of when the customer joins the fund. And Daily Rolling analysis provides a sufficient number of samples required to obtain statistics on the operating period. 

As before the rolling experiment, the Markowitz \cite{markowitz1952portfolio} yields higher returns than the other models, but with much higher volatility. This indicates that Markowitz does not have stable asset management.

In the case of risk budgeting \cite{pearson2011risk}, it can be seen that although volatility is low compared to other models, the return is very low compared to other models. In the case of daily rolling, it can be seen that the asset allocation ratio calculated by the risk budget \cite{pearson2011risk} is concentrated in bonds.

For Daily Rolling, Equal Weight has a lower average annual return than Markowitz, but much lower average annual volatility.This means that asset management performance of Equal weight is not as bad as that of Markowitz. And while equal weight have greater average annualized variability than risk budgeting models, Equal Weight have higher average annualized returns. 

As in the case of Not Daily Rolling, we can confirm that our model is performing well compared to the benchmarks. This shows that our model is not limited to a specific asset class and shows satisfactory performance in various asset classes.

\begin{table*}[h]
\centering
\addtolength{\leftskip} {-2cm}
\addtolength{\rightskip}{-2cm}
\caption{Experiment Result, Rolling}
\label{t5}
\resizebox{16cm}{!}{
\begin{tabular}{|c|c|c|c|c|c|c|c|c|c|c|c|c|c|c|c|c|}
\hline\hline

 &
\multicolumn{4}{|c|}{Markowitz} &
\multicolumn{4}{|c|}{Risk Budgeting} &
\multicolumn{4}{|c|}{Equal Weight} &
\multicolumn{4}{|c|}{Model} \\

\hline
Period & R & Sigma & MDD & Sharpe & R & Sigma & MDD & Sharpe & R & Sigma & MDD & Sharpe & R & Sigma & MDD & Sharpe \\

\hline
(1) 
& 12.83 $\%$ & 35.49$\%$ & -33.25$\%$ & 0.36 
& 3.64$\%$ & 4.07$\%$ & -4.28$\%$ & 0.89 
& 6.39$\%$ &  8.60$\%$ & -11.27$\%$ & 0.74 
& 5.56$\%$ & 5.99$\%$ & -6.82$\%$ &  0.92 \\

\hline
(2) 
& 9.56$\%$ & 17.49$\%$ & -18.77$\%$ & 0.54 
& 2.80$\%$ & 2.50$\%$ & -3.24$\%$ & 1.12 
& 4.52$\%$ &  6.64$\%$ & -7.78$\%$ & 0.68 
& 4.89$\%$ & 4.78$\%$ & -4.81$\%$ &  1.02 \\

\hline
(3) 
& 13.69$\%$ & 14.72$\%$ & -13.78$\%$ & 0.93 
& 2.10$\%$ & 2.32$\%$ & -4.42$\%$ & 0.9 
& 4.40$\%$ &  4.91$\%$ & -7.21$\%$ & 0.89 
& 3.56$\%$ & 4.31$\%$ & -6.09$\%$ &  0.82 \\

\hline
(4) 
& 6.14$\%$ & 21.66$\%$ & -26.97$\%$ & 0.28 
& 2.19$\%$ & 2.09$\%$ & -3.63$\%$ & 1.04 
& 3.46$\%$ &  5.42$\%$ & -8.77$\%$ & 0.63 
& 2.59$\%$ & 4.28$\%$ & -7.20$\%$ &  0.6 \\

\hline
(5) 
& 8.39$\%$ & 15.45$\%$ & -18.97$\%$ & 0.54 
& 0.61$\%$ & 2.02$\%$ & -3.74$\%$ & 0.3 
& 2.44$\%$ &  4.42$\%$ & -8.01$\%$ & 0.55 
& 2.47$\%$ & 3.73$\%$ & -6.10$\%$ &  0.66 \\

\hline
(6) 
& 10.76$\%$ & 24.46$\%$ & -33.10$\%$ & 0.44 
& 2.05$\%$ & 3.39$\%$ & -7.42$\%$ & 0.6 
& 5.30$\%$ &  7.87$\%$ & -14.92$\%$ & 0.67 
& 5.04$\%$ & 6.29$\%$ & -11.93$\%$ &  0.8 \\

\hline
Mean 
& 10.23$\%$ & 21.55$\%$ & -24.14$\%$ & 0.47 
& 2.23$\%$ & 2.73$\%$ & -4.46$\%$ & 0.81 
& 4.42$\%$ &  6.31$\%$ & -9.66$\%$ & 0.70 
& 4.02$\%$ & 4.90$\%$ & -7.16$\%$ & 0.82 \\

\hline
\hline
\end{tabular}
}
\end{table*}

\section{Conclusion}
In this paper, we avoid over-optimization of the model using Monte Carlo simulation data, and consider the statistical structure of the financial market in model learning. We also consider the characteristics of financial markets and portfolio management in the reinforcement learning agent state design. These method show performance that outperforms benchmarks in various test intervals. The method proposed in this paper has the advantage that it can be applied to actual asset management and can be applied to various asset classes.

\bibliographystyle{ACM-Reference-Format}
\bibliography{bib_example}

\appendix

\end{document}